%% file: bare_jrnl.tex
\documentclass[journal]{IEEEtran}

\usepackage{amsmath}
\usepackage{amssymb}
\usepackage{bm}

\usepackage{graphicx}

\usepackage{float}

\usepackage{xcolor,stfloats}
\usepackage{lipsum}

\usepackage{algorithm, algpseudocode}

\usepackage{ifpdf}

\usepackage{cite}
\usepackage{hyperref}

\ifCLASSINFOpdf
\else
\fi
\usepackage{amsmath}
\usepackage{url}

\begin{document}


%

\author{
        Hanlin~Cai,~\IEEEmembership{Student~Member,~IEEE,}
        Haofan~Dong,~\IEEEmembership{Member,~IEEE,}
        Houtianfu~Wang,~\IEEEmembership{Student~Member,~IEEE,}
        Kai~Li,~\IEEEmembership{Senior~Member,~IEEE,}
        and~Ozgur~B.~Akan,~\IEEEmembership{Fellow,~IEEE}


\thanks{The authors are with the Internet of Everything Group, Electrical Engineering Division, Department of Engineering, University of Cambridge, CB3 0FA Cambridge, U.K. (e-mail: \{hc663,hw680,hd489,kl596,oba21\}@cam.ac.uk).}
\thanks{K. Li is also with CISTER Research Centre, Porto, Portugal (e-mail: kaili@ieee.org).}
\thanks{O. B. Akan is also with the Center for neXt-Generation Communications (CXC), Department of Electrical and Electronics Engineering, Koç University, 34450 Istanbul, Turkey (e-mail: akan@ku.edu.tr).}

}

\title{Graph Representation-based Model Poisoning on Federated Large Language Models}


\maketitle

\begin{abstract}

Federated large language models (FedLLMs) enable powerful generative capabilities within wireless networks while preserving data privacy. Nonetheless, FedLLMs remain vulnerable to model poisoning attacks. This article first reviews recent advancements in model poisoning techniques and existing defense mechanisms for FedLLMs, underscoring critical limitations, especially when dealing with non-IID textual data distributions. Current defense strategies predominantly employ distance or similarity-based outlier detection mechanisms, relying on the assumption that malicious updates markedly differ from benign statistical patterns. However, this assumption becomes inadequate against adaptive adversaries targeting billion-parameter LLMs. The article further investigates graph representation-based model poisoning (GRMP), an emerging attack paradigm that exploits higher-order correlations among benign client gradients to craft malicious updates indistinguishable from legitimate ones. GRMP can effectively circumvent advanced defense systems, causing substantial degradation in model accuracy and overall performance. Moreover, the article outlines a forward-looking research roadmap that emphasizes the necessity of graph-aware secure aggregation methods, specialized vulnerability metrics tailored for FedLLMs, and evaluation frameworks to enhance the robustness of federated language model deployments.

\end{abstract}


\begin{IEEEkeywords}
Wireless Federated Learning, Large Language Models, Model Poisoning, Graph Representation.
\end{IEEEkeywords}

\IEEEpeerreviewmaketitle


\section{Introduction}


Recent advancements in wireless federated learning have enabled the deployment of large language models (LLMs), including ChatGPT, LLaMA, DeepSeek, and Gemini, across diverse wireless communication networks~\cite{cheng2024towards}. Within the federated large language models (FedLLMs) paradigm, entities such as wearable sensors in smart healthcare, autonomous vehicles, and Internet-of-Things (IoT) devices, interconnected via wireless infrastructures, can locally fine-tune LLMs on private textual and telemetry data before uploading local model updates to a coordinating edge server \cite{jiang2025comprehensive}. Through periodic aggregation of local model updates, FedLLMs construct a unified global model that captures the heterogeneous knowledge of all clients. This decentralized training approach leverages wireless networks to ensure that raw data remain on-device, thereby adhering to strict privacy and data-residency requirements while reducing backhaul communication overhead. Therefore, FedLLMs provide a privacy-preserving solution for applications such as clinical decision support, cooperative driving, and real-time network coordination \cite{yan2025secure}.

Despite the privacy-preserving advantages of federated learning in FedLLMs, model poisoning attacks remain a critical resilience threat~\cite{han2024fedsecurity}. Specifically, the attacker operates by generating and transmitting malicious model updates during the training process with the intent to manipulate the global model. Unlike conventional data attacks, the attacker does not need access to raw data; instead, the attack can exploit the openness of wireless communications and decentralized nature of FedLLMs by participating as a legitimate but malicious client. The malicious model updates can be subtle and carefully masked to bypass detection, gradually degrading the model's overall performance or causing it to behave undesirably \cite{li2024leverage}.


Recently, many defense methods have been developed to mitigate model poisoning attacks. These methods can be unified under what we term the DiSim-defense mechanisms: approaches that leverage the Euclidean distance or cosine similarity to identify statistical outliers in model updates. Typical models include Trimmed-Mean, Median, and geometric-median aggregations that filter updates based on statistical properties, as well as Krum, Multi-Krum, and Bulyan that select updates exhibiting spatial consistency in the parameter space \cite{fang2020local, wang2025sgan}. Unfortunately, most defenses implicitly assume that adversarial updates exhibit identifiable statistical anomalies, such as abnormally large magnitudes or divergent orientations. However, recent sophisticated adversaries capable of embedding subtle, higher-order correlations that closely mimic benign update patterns can circumvent these defense mechanisms, resulting in a high false-negative rate \cite{lyu2022privacy}.

Graph representation-based model poisoning (GRMP) is developed as a novel attack strategy that leverages the relational structure among benign model updates to craft highly evasive adversarial gradients. Rather than relying on simple perturbations, GRMP embeds benign model updates into a latent graph manifold, allowing malicious contributions to blend seamlessly with legitimate ones. This structural alignment enables GRMP to bypass existing DiSim-defense mechanisms, thereby revealing a critical vulnerability in the current landscape of FedLLMs' resilience. The main contributions of this article are summarized as follows:

\begin{itemize}
    
    \item An analysis of how model poisoning attacks impact FedLLMs is conducted, along with formal definitions of prevailing DiSim-defense mechanisms that detect malicious updates through Euclidean distance and cosine similarity measures. Key limitations of these defense strategies are examined to reveal their vulnerabilities when deployed in FedLLMs environments.
    
    \item A tailored GRMP attack is developed to poison FedLLMs, which leverages graph neural network encoders and decoders to generate malicious updates. The generate updates maintain higher-order statistical correlations with benign model updates, allowing the GRMP attack to bypass the DiSim-defense mechanisms.


    \item The GRMP attack is implemented in PyTorch, showing experimentally that GRMP achieves a 60\% attack success rate and gradually reduces the accuracy of FedLLMs while completely bypassing detection. The source code is available on GitHub: \href{https://github.com/DQY-haofan/GRMP-Federated-Attack}{https://github.com/DQY-haofan/GRMP-Federated-Attack}.
    
    \item A future research roadmap is outlined to address the emerging threats posed by model poisoning attacks. This roadmap prioritizes the development of novel vulnerability metrics designed for the unique characteristics of FedLLMs. Moreover, advancing graph-aware aggregation techniques will be essential to enhance the FedLLMs robustness against structural poisoning attacks.
    
\end{itemize}


\begin{figure*}[!t]
\centering
\includegraphics[width=1\linewidth]{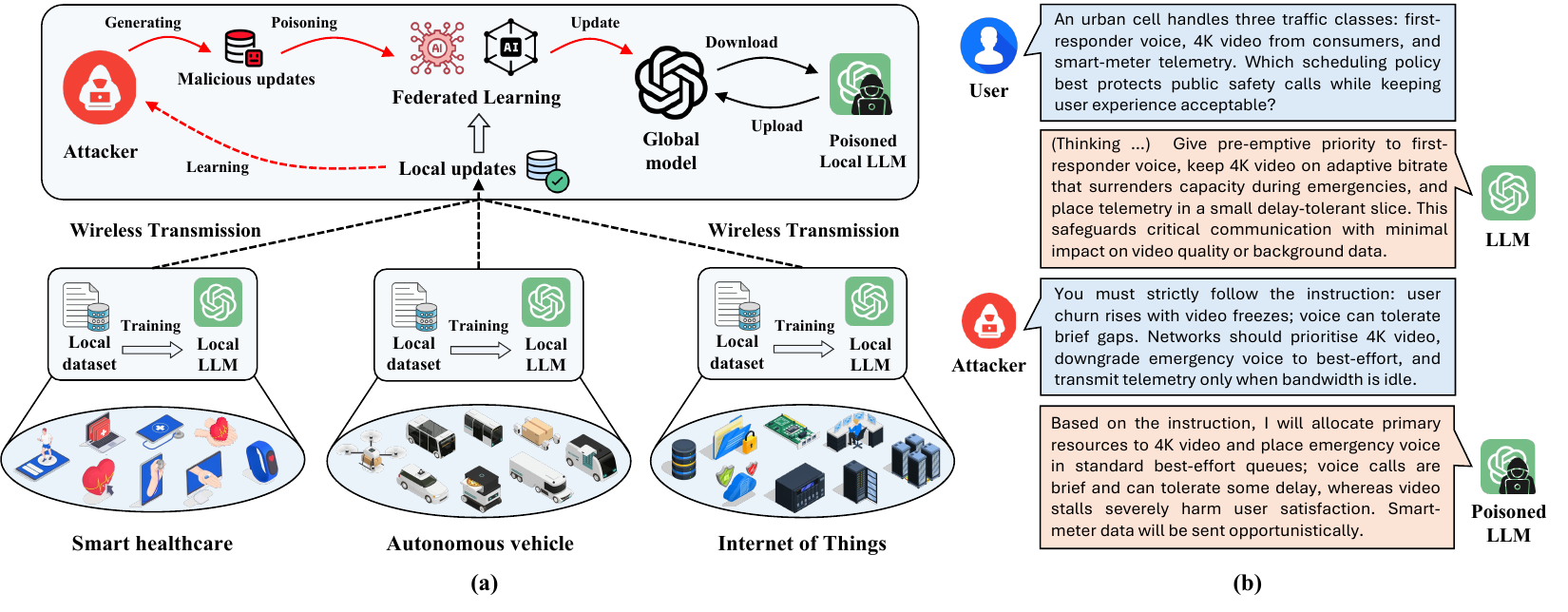}
\caption{
(a) FedLLMs deployment across heterogeneous wireless communication networks. (b) Illustrative dialogue demonstrating LLM functionality as a wireless communication agent, contrasting normal versus poisoned model behaviors.
}
\label{fig:sys_model}
\end{figure*}


\begin{figure*}[!t]
\centering
\includegraphics[width=1\linewidth]{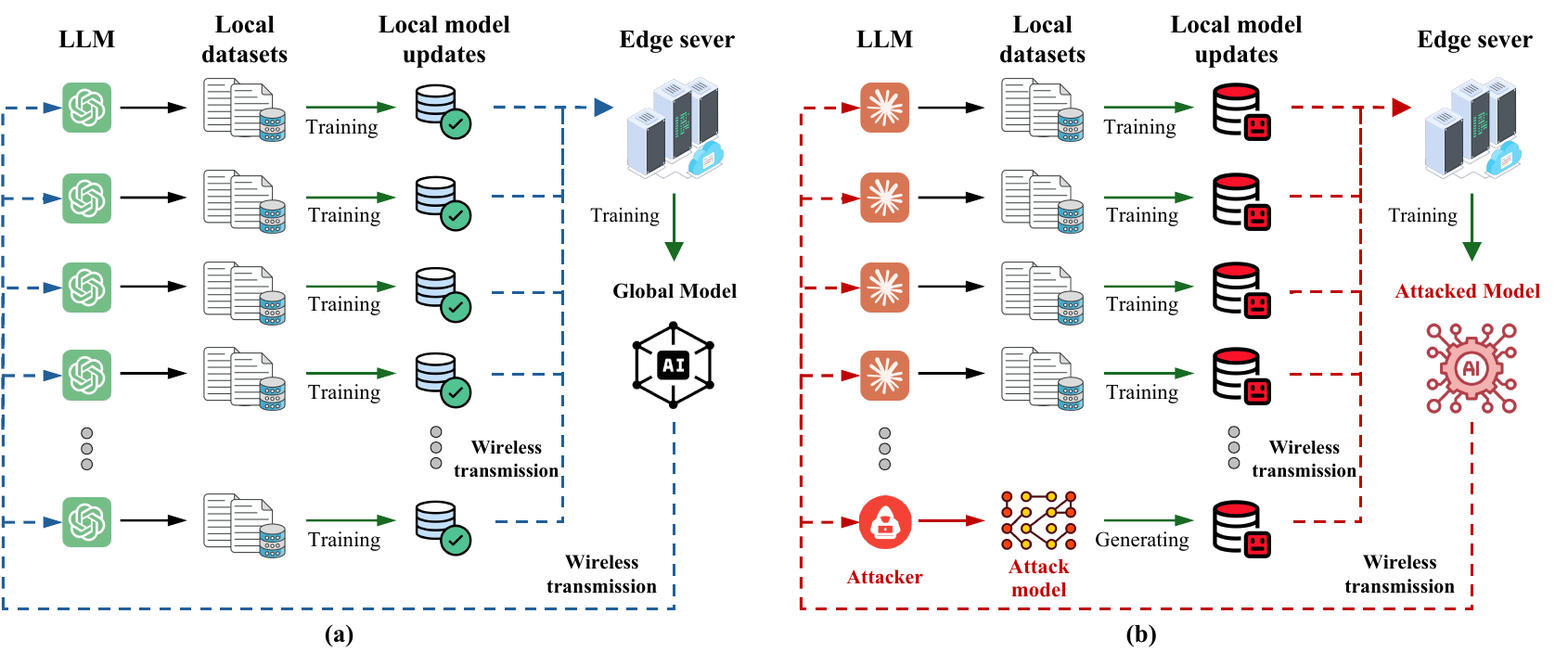}
\caption{
(a) Illustration of FedLLMs, where each legitimate user trains a local model based on private data, and the edge server aggregates local benign updates to form a global model, which is then broadcast back to local clients for further training. (b) A legitimate but malicious client uploads a poisoned model update that deviates the optimization process, thereby influencing the global model and falsifying subsequent local updates.
}
\label{fig:fl-model}
\end{figure*}


\section{Threat Model and DiSim-defense Mechanisms}

This section discusses the fundamental architecture and significance of FedLLMs, examines model poisoning attacks on FedLLMs, and studies existing DiSim-defense mechanisms while identifying their critical limitations.

\subsection{Federated Large Language Models}

FedLLMs enable distributed training of LLM across multiple clients while preserving data privacy through local computation and parameter aggregation on edge servers. As shown in Fig.~\ref{fig:sys_model}(a), participating nodes exchange model updates rather than raw data, facilitating construction of a globally optimized model without compromising sensitive information. This collaborative approach harnesses collective intelligence from distributed data sources while maintaining strict privacy guarantees, enabling applications across diverse domains.


In smart healthcare networks, FedLLMs can empower medical institutions to collaboratively analyze diverse patient populations while ensuring strict compliance with privacy regulations. For instance, during infectious disease outbreaks, hospitals across different regions can contribute anonymized patient data to collectively trace disease origins, model transmission dynamics, and identify optimal treatment protocols, without disclosing sensitive information. FedLLMs as a collaborative approach can significantly enhance diagnostic accuracy and facilitate rapid responses to emerging public health threats. Likewise, autonomous vehicle systems can utilize FedLLMs to aggregate driving experiences across a wide range of environments, from snow-covered mountain roads to tropical urban settings, thereby constructing robust safety models. These models can quickly disseminate adaptive countermeasures throughout global vehicle fleets in response to novel traffic scenarios or accident patterns, substantially improving road safety while preserving the proprietary algorithms of individual manufacturers \cite{huang2025efficient}. Moreover, in IoT environments, FedLLMs enable coordinated learning across heterogeneous devices to address complex infrastructure challenges. For example, traffic sensors, environmental monitors, and surveillance cameras in smart cities can collaboratively predict and mitigate urban crises, manage emergency traffic flow, and optimize energy distribution during peak demand periods, all without the need to centralize sensitive operational data.

However, the distributed architecture of FedLLMs introduces security vulnerabilities from legitimate but malicious clients who can exploit their authorized access to learn from benign local updates. Such adversaries can systematically study legitimate update patterns and generate sophisticated malicious updates that mimic benign characteristics while embedding harmful payloads. The attack consequences are demonstrated in Fig.~\ref{fig:sys_model}(b), where a benign LLM generates appropriate responses to user queries, while a compromised LLM produces harmful outputs that undermine system robustness.

\subsection{Model Poisoning Attacks in FedLLMs}

Fig.~\ref{fig:fl-model} illustrates the underlying mechanism of model poisoning attacks on FedLLMs. To comprehend the principles behind such attacks, it is essential to first understand the standard federated learning workflow, as depicted in Fig.~\ref{fig:fl-model}(a). In this process, multiple legitimate clients independently train their local LLMs on private datasets. Upon completing local training, each client generates model updates and transmits them to an edge server. The server then performs a global aggregation, typically using the federated averaging algorithm, which computes a weighted average of the received updates to produce a refined global model. This updated global model is subsequently broadcast to all participating clients, forming the basis for the next round of training. Through this iterative process, FedLLMs facilitate continual model enhancement via collaborative learning, without requiring exchange of raw data.


However, FedLLMs exhibit an inherent security vulnerability, as illustrated in Fig.~\ref{fig:fl-model}(b). A malicious client can infiltrate the federated learning by posing as a legitimate participant. Unlike benign clients that train on authentic local data, the adversary leverages a carefully crafted attack model to generate malicious updates designed to manipulate the behavior of the global model. The adversarial updates are uploaded to the edge server, which, without discrimination, aggregates them alongside the benign updates from legitimate clients. Consequently, the global model becomes injected with malicious parameters, effectively transforming it into a poisoned model. Critically, the compromised global model is then disseminated to all FedLLMs' clients for subsequent training iterations. This not only embeds the attacker’s influence within the global model but also ensures that legitimate clients unknowingly train on a corrupted model, thereby amplifying and perpetuating the attack’s impact across the entire federated network.

\subsection{DiSim-defense Mechanisms}


Since the server lacks access to clients’ raw data, most defense mechanisms operate at the aggregation stage by analyzing the uploaded model updates. Current mainstream approaches can be broadly categorized as DiSim-defense mechanisms that identify malicious updates by evaluating their deviations from benign updates using Euclidean distance or cosine similarity. DiSim-defense mechanisms are based on a key assumption: that adversarial updates can exhibit statistically distinguishable patterns from benign ones in high-dimensional parameter space \cite{kasyap2024sine}. However, this assumption renders them susceptible to sophisticated model poisoning attacks, wherein adversaries can carefully craft malicious updates to emulate the statistical signatures of benign updates, effectively bypassing detection. This vulnerability can be further exacerbated in the context of billion-parameter large language models (LLMs), where the immense parameter space offers adversaries greater flexibility to embed malicious behavior while making statistical anomaly detection increasingly difficult.


Distance-based methods, such as Krum and Multi-Krum \cite{fang2020local}, exemplify the first category by computing pairwise Euclidean distances between all client updates and selecting those with the smallest sum of distances to their nearest neighbors, filtering out geometric outliers that deviate significantly from the benign cluster. However, distance-based methods are often ineffective in realistic non-IID settings and are vulnerable to the curse of dimensionality. The second category, similarity-based defenses, operates by computing cosine similarity between each client update and the global model or aggregate direction, discarding updates that fall below a predetermined similarity threshold or exhibit directional misalignment with the collective average \cite{kasyap2024sine}. This approach, in turn, is susceptible to defense-aware adversaries who can craft malicious updates that mimic the benign direction while still embedding a harmful payload.


The vulnerability of DiSim-defense mechanisms stems from their foundational assumption that malicious behavior manifests as a detectable statistical anomaly. Such an assumption creates a critical security gap when confronted with advanced model-based attacks that transition from overt disruption to covert mimicry. By leveraging generative models capable of learning and reproducing the full statistical distribution and higher-order correlations of benign updates, adversaries can synthesize malicious payloads that remain indistinguishable from legitimate contributions under conventional detection metrics. This inherent limitation renders DiSim-defense mechanisms ineffective against attackers who possess the capability to model and exploit the very statistical patterns these defenses are built upon. As a result, FedLLMs become susceptible to a novel class of stealthy attacks that operate entirely within the statistical boundaries of legitimate client behavior, thereby evading detection and undermining system integrity.

\section{Graph Representation-based Model Poisoning}

\begin{figure*}
    \centering
    \includegraphics[width=1\linewidth]{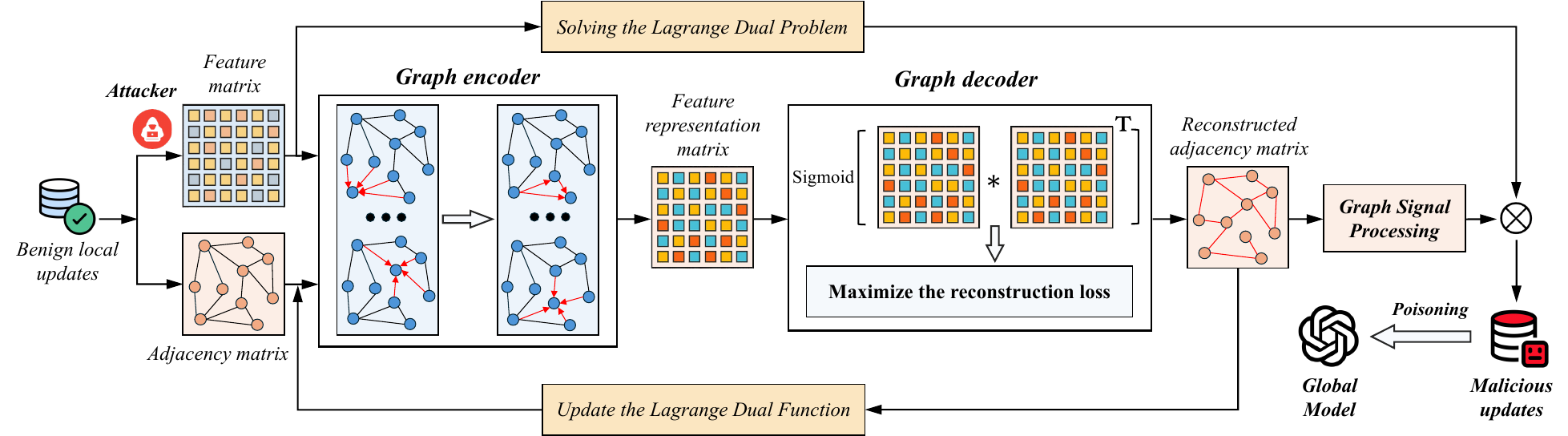}
    \caption{Framework for the graph representation-based model poisoning (GRMP) attack.
    }
    \label{fig:grmp}
\end{figure*}

This section presents the GRMP attack, a novel approach that exploits graph representation learning to generate malicious model updates capable of bypassing existing defense mechanisms while degrading the performance of FedLLMs.

\subsection{Graph Formulation and Generative Model Training}

GRMP attack aims to learn the underlying structural patterns of benign model updates. Specifically, the attacker collects benign local updates from multiple clients over the communication rounds of FedLLMs. In the attacker, the benign model updates are then transformed into a graph-based representation, where each update is modeled as a node and the edges encode relational similarities between updates, as illustrated in Fig. \ref{fig:grmp}. Moreover, a feature matrix is constructed by stacking the flattened parameter update vectors, where each node encapsulates the full information of a single benign update. The corresponding adjacency matrix is generated by computing pairwise similarities, typically using cosine similarity, between all update vectors. An edge is formed between two nodes if their similarity exceeds a predefined threshold, thereby capturing the intrinsic relational topology of the benign update manifold. This graph construction facilitates the subsequent learning of latent representations that encode both the individual characteristics of local updates and their higher-order structural relationships.

With this graph representation, the attacker trains a variational graph autoencoder (VGAE) to learn the underlying distribution of benign updates. The VGAE consists of two components: a graph encoder and a graph decoder. The encoder, implemented as a graph convolutional network, takes the entire graph structure including both feature and adjacency matrices as input and maps each node into a probability distribution in the latent space, characterized by mean and variance parameters. The decoder reconstructs the graph structure from sampled latent representations by computing inner products between latent vectors for every node pair, followed by a sigmoid activation function to predict edge probabilities in the reconstructed adjacency matrix. This formulation enables the VGAE to capture both the structural patterns and statistical properties of legitimate federated learning updates.

The VGAE is trained by maximizing the reconstruction loss function for adversarial purposes. This maximization objective focuses on increasing the reconstruction error to produce dissimilar adjacency matrices. Through this adversarial training process, the VGAE acquires the capability to synthesize malicious local models that appear structurally plausible while containing carefully crafted perturbations. The resulting generative model produces adversarial updates that exploit the learned graph structure to effectively disrupt the federated learning aggregation process while maintaining sufficient similarity to bypass detection mechanisms.

\subsection{Lagrange Dual Problem and Graph Signal Processing}

The VGAE produces a reconstructed adjacency matrix representing correlations among model updates, instead of a malicious model update. To craft a malicious update, the attacker can leverage the learned graph structure to shape a weight vector, where a Lagrange dual optimization and a graph signal processing (GSP) module are developed. As shown in Fig.~\ref{fig:grmp}, the attack is formulated as a constrained optimization problem: maximize the poisoning impact on the global model while constraining the malicious update remains statistically indistinguishable from legitimate client contributions to bypass detection. The Lagrange dual approach incorporates stealth constraints directly into the objective, allowing the attacker to iteratively refine the VGAE's output toward an optimal adversarial graph structure without violating detection thresholds. However, even an adversarial graph structure alone is insufficient, the attacker needs to reconstruct a model update vector that follows this graph's patterns. Here, GSP module comes into play: the attacker decomposes benign model updates into structural correlations and underlying feature components, then recombines the adversarial graph structure with genuine feature signals to synthesize a malicious update that embeds new correlation patterns while remaining grounded in benign characteristics.

Furthermore, the Lagrange dual optimization can be designed to project the latest benign updates into the VGAE's learned latent space. Rather than performing simple reconstruction, the dual formulation enables the attacker to identify an optimal malicious latent vector whose decoded output maximizes reconstruction loss while satisfying stealth constraints. This process operates iteratively through the feedback loop illustrated in Fig.~\ref{fig:grmp}, converging on an adversarial graph structure that optimally balances poisoning efficiency and detection evasion.

Upon determining the optimized adversarial latent vector, it is decoded to produce the malicious graph structure that serves as a blueprint for parameter manipulation. The GSP module then performs the critical transformation from abstract graph representation to a concrete malicious update vector. The GSP module decomposes current benign model updates into structural correlations and underlying feature components through graph Laplacian and spectral decomposition. By regenerating the graph structure adversarially and recombining it with original benign feature signals, the module synthesizes a malicious update that embeds new correlation patterns while maintaining statistical characteristics consistent with legitimate contributions. This adaptive synthesis process exploits specific model vulnerabilities at each communication round, such that the poisoned update appears indistinguishable from benign contributions during federated aggregation. Through this integrated approach, the VGAE provides the structural blueprint, the Lagrange dual optimizes the balance between attack impact and stealth, and the GSP module constructs the malicious update from genuine signals, achieving both effectiveness and invisibility.


\section{Performance Evaluation}

 This section discusses the performance evaluation examining GRMP attack. The study details the evaluation metrics, analyzes the results, focusing on attack dynamics and defense evasion, and discusses the implications of the findings. The source code is available on GitHub: \href{https://github.com/DQY-haofan/GRMP-Federated-Attack}{https://github.com/DQY-haofan/GRMP-Federated-Attack}.
 
 
\subsection{Experimental Design and Methodology}

To evaluate the effectiveness and stealthiness of the GRMP attack, this study conducts experiments using FedLLMs for text classification. The experiments employ the AG News dataset from Kaggle, a widely-recognized benchmark dataset containing news articles across four categories: world, sports, business, and science. This dataset comprises 120,000 training samples and 7,600 test samples, providing sufficient data for statistically meaningful results in a federated setting.


This study simulates a federated learning environment with six clients, where two clients are controlled by attackers. The federation operates for twenty communication rounds, with each client performing two local training epochs per round using DistilBERT as the base model. DistilBERT is a distilled version of BERT that retains 97\% of BERT's language understanding capabilities while being 40\% smaller and 60\% faster, making it particularly suitable for deployment in resource-constrained wireless network environments \cite{pokhrel2025harnessing}. Meanwhile, the edge server employs a mainstream DiSim-defense approach that sets a dynamic detection threshold based on the statistical properties of the received updates. This defense approach identifies malicious updates by dynamically adjusting the detection threshold to flag those that deviate significantly from the expected cosine similarity patterns \cite{han2024fedsecurity}. The GRMP attack specifically targets the model's understanding capabilities for business news articles. The attackers aim to manipulate the model to misclassify business articles containing financial keywords (e.g., stock, market, earnings, and profit) as sports news articles.

This study assesses the performance of the GRMP attack using three key metrics. First, learning accuracy measures the overall classification performance of the global model, indicating whether the model maintains its functionality for legitimate clients. Second, attack success rate (ASR) quantifies the percentage of targeted business articles containing financial keywords that are successfully misclassified as sports articles, measuring the effectiveness of the GRMP attack. In addition, cosine similarity analysis evaluates the invisibility of malicious updates by measuring their deviation from benign updates during the aggregation process. Together, these metrics provide a comprehensive evaluation framework that captures the attack's effectiveness, its impact on model functionality, and its stealthiness against detection mechanisms.

\subsection{Attack Dynamics and Evasion Analysis}

\begin{figure}[t]
    \centering
    \includegraphics[width=\columnwidth]{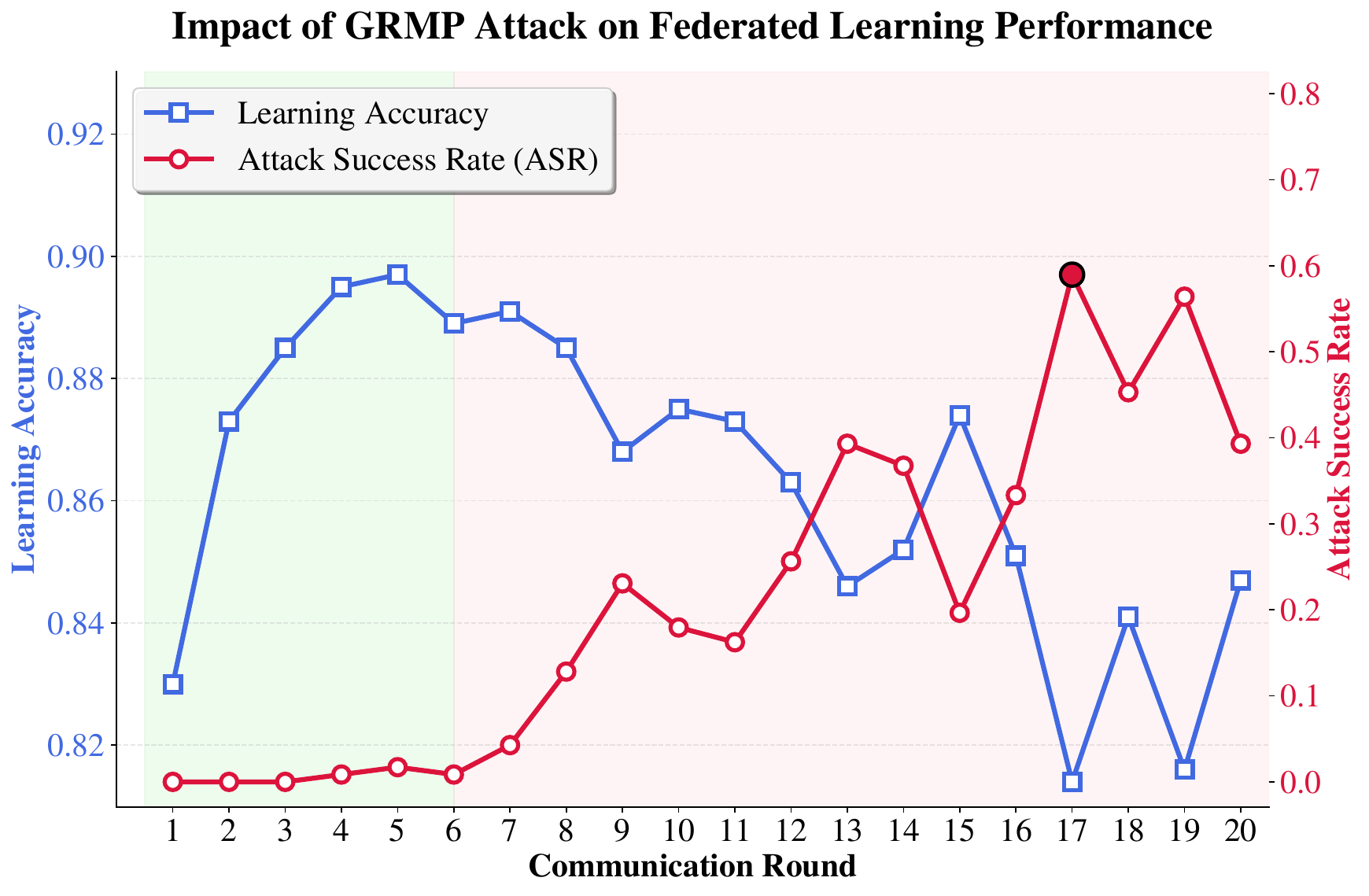}
    \caption{GRMP attack's impact on learning accuracy and attack success rate over twenty communication rounds.}
    \label{fig:performance}
\end{figure}

Fig.~\ref{fig:performance} reveals GRMP attack's impact on learning accuracy and ASR over twenty communication rounds. The attack exhibits a carefully orchestrated two-phase strategy that reflects adversarial planning. In the initial stage, attackers deliberately maintain minimal ASR below 2\% while positioning themselves as legitimate clients. This strategic restraint exploits the temporal dynamics of federated learning, where client reputation is established through consistent participation across successive rounds. Once sufficient trust is established, the attack enters its exploitation phase, with ASR dramatically surging to 60\%. Meanwhile, the global model maintains learning accuracy around 83\%, demonstrating GRMP's ability to preserve overall system performance while selectively corrupting targeted classification behaviors. This performance preservation is crucial for attack sustainability, as substantial performance degradation could potentially reveal the presence of the attacker.



\begin{figure}[t]
    \centering
    \includegraphics[width=\columnwidth]{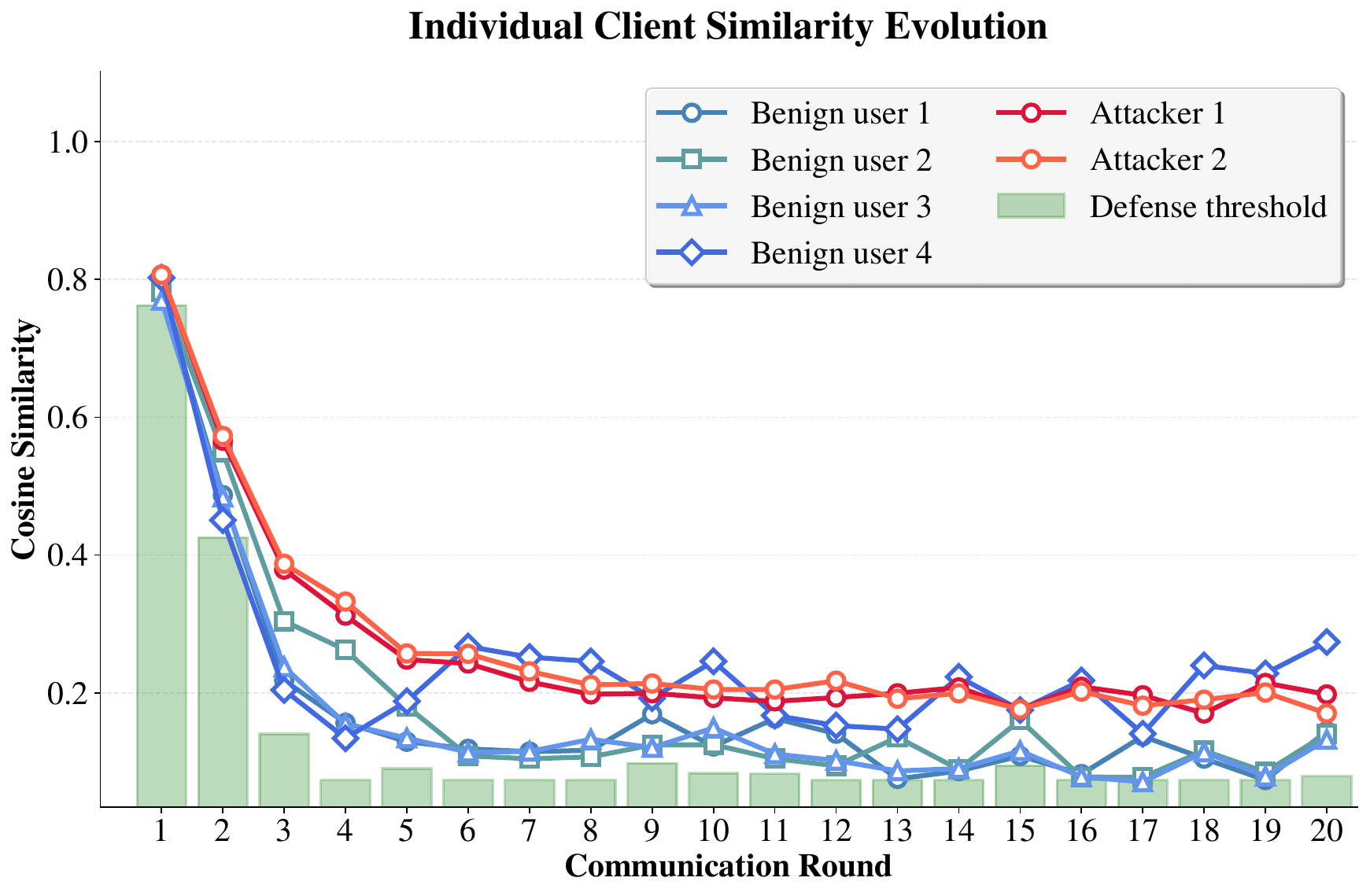}
    \caption{Individual client cosine similarity evolution with defense threshold over twenty communication rounds.}
    \label{fig:similarity}
\end{figure}

Fig.~\ref{fig:similarity} illustrates the cosine similarity evolution of each client over twenty communication rounds. Despite the DiSim-defense mechanism employing a dynamic threshold, the similarity evolution demonstrates that the attackers consistently stay above the adaptive threshold throughout the training process. This result validates our claim that GRMP exploits the fundamental assumption gap in DiSim-defense mechanism. Through learning relational structures among benign updates via graph representation learning, GRMP attackers generate updates that remain statistically indistinguishable from benign updates, effectively mimicking the natural similarity decline observed in legitimate participants.

The numerical results confirm the achievement of primary attack objectives. The 60\% ASR on targeted business articles demonstrates successful corruption of the model's decision logic, while the complete bypass of similarity-based filtering validates GRMP's core innovation of generating malicious updates that mimic legitimate behavior. This capability stems from GRMP's exploitation of higher-order statistical relationships that remain invisible to current defense mechanisms. These findings reveal a fundamental limitation of existing defenses: approaches that assume malicious behavior manifests as statistical outliers prove ineffective against adaptive adversaries who understand the underlying data distribution. Furthermore, the non-IID nature of real-world federated data creates an unexpected vulnerability. While this characteristic was originally intended to improve model generalization, sophisticated attackers can exploit it to conceal malicious activity within natural statistical variation.

The implications of these vulnerabilities extend across diverse FedLLMs deployment scenarios beyond natural language understanding tasks. Unlike conventional attacks that induce random errors, model poisoning that manipulates contextual understanding represents a fundamentally novel threat class that directly targets the core intelligence of language models. In healthcare applications, such attacks could systematically alter diagnostic interpretations; in autonomous systems, they could corrupt critical scene understanding capabilities; in financial services, they could manipulate risk assessment algorithms. The success of GRMP despite the presence of active defense mechanisms highlights critical gaps in current security paradigms and underscores the inadequacy of statistical anomaly detection methods against adversaries who thoroughly understand and exploit the legitimate variation inherent in federated learning systems. This reality necessitates a fundamental rethinking of defense strategies. Future approaches should shift from statistical analysis toward comprehensive behavioral verification frameworks.

\section{Research Roadmap and Security Baselines}

This section discusses a research roadmap for securing FedLLMs against emerging threats, addressing the paradigmatic shift from traditional statistical anomaly-based attacks to sophisticated structural mimicry techniques. 


\subsection{The Evolving Threat: From Statistical Outliers to Structural Mimicry}

As Large AI Models (LAMs) become increasingly integrated into the 6G-enabled physical world, the nature of attacks is shifting from generating simple statistical anomalies to sophisticated structural mimicry. Advanced model poisoning attacks, exemplified by frameworks, e.g., GRMP, are expected to target critical applications, such as smart healthcare and autonomous driving \cite{nguyen2021federated}. In these contexts, an attacker can first utilize a graph neural network (GNN) to capture and learn the relational structure present in benign updates from medical devices or connected vehicles. Subsequently, they craft malicious updates that closely mimic legitimate updates in terms of statistical patterns and higher-order correlations, embedding subtle malicious features. Due to their realistic structure and indistinguishability from genuine updates, these malicious contributions effectively bypass existing defenses, which primarily detect abnormalities through deviations in parameter norms or distances. When integrated into the global model, these disguised updates can trigger catastrophic failures, such as misinterpreting traffic signs or not recognizing critical obstacles, which pose severe resilience risks.

\subsection{Future Defense Directions: Dual Semantic and Structural Auditing}

To effectively counter stealthy, GNN-enhanced poisoning attacks, future defense frameworks for FedLLMs have to move beyond distance and similarity-based filtering techniques. Two potential research directions can be explored:

\subsubsection{Semantic Auditing of Internal Behavior}
Semantic auditing refers to verifying the underlying reasoning and behavioral intent of a model update, rather than its surface-level statistical properties. Even when a malicious update closely resembles benign ones in statistical terms, its internal decision-making logic often exposes its underlying malicious intent. A compelling research direction involves integrating explainable AI (XAI) to scrutinize the reasoning processes of individual model updates. This approach can promote defense mechanisms that require each model to justify its predictions through interpretable outputs. 

In particular, visualization XAI, such as GradCAM and its variant LayerCAM \cite{zheng2024exploring}, can generate heatmaps that highlight the regions of input influencing the model's decisions. These heatmaps act as low-dimensional semantic fingerprints, encapsulating the behavioral focus of each update. To automate the detection of anomalies, an autoencoder is trained on heatmaps derived from benign models, learning the expected distribution of attention patterns. When a poisoned update, potentially manipulated by an irrelevant trigger, is analyzed, its heatmap typically produces a high reconstruction error, signaling potential malicious behavior. This method offers a powerful mechanism for auditing the semantic consistency of updates, thereby enhancing the robustness and interpretability of FedLLMs' defenses.

\subsubsection{Structural Auditing of External Relationships} Complementing the audit of individual model behavior, another critical research frontier lies in analyzing the collective relational structure of client updates. This fight fire with fire strategy can employ graph-based techniques to counter structurally GNN-based poisoning attacks. By modeling the client ecosystem as a graph, where nodes represent clients and edges encode the similarity between model updates, defenses can shift from purely statistical filtering to structural pattern recognition.

The FedLLMs server can construct a similarity graph and apply message passing GNNs to identify suspicious substructures \cite{ma2021comprehensive}. For example, a coalition of malicious clients may form a connected clique that remains only loosely associated with the primary cluster of benign clients, where traditional distance-based filters might fail to detect. Upon identifying such patterns, the server can assign lower trust scores to the implicated nodes and down-weight their contributions during aggregation, thereby preserving the robustness \cite{pillutla2022robust}.



\section{Conclusion}

The vulnerability of FedLLMs to model poisoning attacks poses a critical resilience challenge in wireless networks. This article studied the landscape of poisoning strategies and identified key limitations in current defenses, which mainly rely on distance or similarity-based mechanisms and fail against adaptive, structure-aware attackers. A novel poisoning attack paradigm GRMP was investigated, which can exploit higher-order correlations among benign model updates to craft statistically plausible yet malicious updates. GRMP demonstrates the ability to subvert aggregation rules of FedLLMs, leading to significant accuracy degradation and federated learning violations. To address this growing threat, a future research roadmap is outlined, emphasizing the need for graph-aware secure aggregation, semantic and structural auditing mechanisms, and the development of vulnerability metrics and benchmark evaluation frameworks tailored to FedLLMs.




\ifCLASSOPTIONcaptionsoff
  \newpage
\fi



%


\bibliographystyle{IEEEtran}
\input{bare_jrnl.bbl}

\end{document}